\documentclass[pra, showpacs, twocolumn, a4paper, nofootinbib]{revtex4}
\usepackage{graphicx}
\usepackage{amsmath, amsfonts, amssymb, bm}
\usepackage{amsmath}
\usepackage{verbatim}
\usepackage{bm}
\usepackage{graphicx,psfrag,bm,times,amsmath,amssymb}

\begin{document}

\title{Comment on "Discrepancies in the resonance-fluorescence spectrum calculated 
with two methods"}
\author{Zbigniew \surname{Ficek}$^{a,b}$}
\email{ficek@physics.uq.edu.au}
\affiliation{$^{a}$Department of Physics, School of Physical Sciences, The University of Queensland, Brisbane, Australia 4072 \\
$^{b}$The National Centre for Mathematics and Physics, KACST, P.O.Box 6086, Riyadh 11442, Saudi Arabia}

\date{\today}

\begin{abstract}
There are two alternative methods used in literature to calculate the incoherent part of the spectrum of light scattered by an atomic system. In the first, one calculates the spectrum of the total light scattered by the system and obtains the incoherent part by subtracting the coherent part. In the second method, one introduces the fluctuation operators and obtains the incoherent part of the spectrum by taking the Fourier transform of the two-time correlation function of the fluctuation operators. These two methods have been recognized for years as two completely equivalent for evaluating the incoherent part of the spectrum.
In a recent paper Qing Xu {\it et al.} [Phys. Rev. A {\bf 78}, 013407 (2008)] have shown that there are discrepancies between the incoherent parts of the stationary spectrum of a three-level $\Lambda$-type system calculated with these two methods.  
The predicted discrepancies can be severe that over a wide rage of the Rabi frequencies and atomic decay rates, the spectrum calculated with the variance method can have negative values. This is obviously unphysical result since the fluorescence spectrum is a positively defined quantity. It represents the frequency distribution of light incoherently scattered by the atomic system. Therefore, the calculated spectrum should be positive for all frequencies independent of values of the Rabi frequencies and the damping rates. 
In this comment, we show that there are no discrepancies between these two methods. The equivalence of these two alternative methods leads to the same incoherent spectra that are positive for all frequencies independent of values of the parameters involved. The analytical analysis is supported by simple numerical calculations.
\end{abstract}


\maketitle

\ \  The incoherent part of the fluorescence spectrum of an atomic system driven by lasers fields is commonly studied with two, often called "limit" and "variance" methods. In the limit method, one calculates the total spectrum from which then the coherent part is subtracted to obtain the incoherent part alone. In the variance method, one define the fluctuation operators and directly obtains the incoherent part of the spectrum by a Fourier transformation of the two-time correlation function of the fluctuation operators.  These methods are generally recognized as completely equivalent and have been used for decades to calculate spectra of the stationary electromagnetic field radiated by variety of atomic systems~\cite{m69,km76,cw76,ws76,ct77,ns90,eb02,ft09}. The variance method has been particularly popular in the calculations of the incoherent spectrum of non-stationary fields due to some difficulties in the treatment of the coherent part of the spectrum~\cite{bf80,cr83,wr89,az91,ff93,ffd96,ff00,ab02}.

\ \ In a recent paper, Xu {\it et al.}~\cite{xhy} have calculated the incoherent spectrum of the fluorescence field emitted by a three-level $\Lambda$ system driven by two laser fields and claim that these two methods produce different results, and the method of the calculation of the spectrum involving the fluctuation operators may even lead to negative values of the spectrum. This statement, however, is incorrect. In this comment we clarify this point, directly showing that both methods lead to the same incoherent part of the spectrum and the negative values of the spectrum simply arise from an error in the Xu {\it et al.}~\cite{xhy} numerical computation. Specifically, the explanation of the source of the discrepancies given by Xu {\it et al.} is based on the general solution of the set of differential equations describing the time evolution of the system. Therefore, it is necessary to show that there are no discrepancies between these alternative methods  since similar discrepancies could be found in other systems described by a similar set of differential equations.

\ \ As it is well known, the spectrum of the fluorescence light emitted by an atomic system is obtained by a Fourier transformation of the two-time correlation function of the emitted electromagnetic field that is
usually expressed in terms of the two-time correlation function of the atomic operators. In the case of the $\Lambda$-type atom considered by Xu {\it et al.}~\cite{xhy}, the spectrum of the fluorescence light emitted on the atomic $|3\rangle - |1\rangle$ transition can be written~as
\begin{eqnarray}
S_{T}(\omega) &=& \gamma_{1}u(\vec{r}){\rm Re}\int_{0}^{\infty} 
d\tau e^{-i(\omega -\omega_{1})\tau} \nonumber \\
&&\times  \lim_{t\rightarrow\infty}\langle \sigma_{31}(t+\tau)\sigma_{13}(t)\rangle ,\label{e2}
\end{eqnarray}
where $\gamma_{1}$ is the spontaneous emission rate of the transition, $u(\vec{r})$ is a geometrical factor corresponding to the radiation pattern of the dipole moment and the expectation values of the atomic dipole operators are evaluated in the rotating frame oscillating with the frequency $\omega_{1}$ of the atomic transition involved.

\ \ Usually the scattered field is composed of a coherent component, corresponding to a field elastically scattered by the source atoms, and an incoherent (noise) component, corresponding to a field produced by fluctuations of the atomic dipoles. Therefore, it is often distinguished between these two contributions to the scattered field by expressing an atomic dipole operator as the sum of its expectation value and its fluctuations so that
\begin{eqnarray}
\sigma_{ij}(t) = \langle\sigma_{ij}(t)\rangle +\Delta \sigma_{ij}(t) ,
\end{eqnarray}
where $\langle\Delta\sigma_{ij}(t)\rangle =0$ by definition. In terms of the fluctuation operators the spectrum is of the form
\begin{eqnarray}
S_{in}(\omega) &=& \gamma_{1}u(\vec{r}){\rm Re}\int_{0}^{\infty} d\tau 
e^{-i(\omega -\omega_{1})\tau} \nonumber \\
&&\times  \lim_{t\rightarrow\infty}
\langle \Delta\sigma_{31}(t+\tau)\Delta\sigma_{13}(t)\rangle  ,\label{e3}
\end{eqnarray}

We illustrate the equivalence of the two methods using three different arguments that correspond to different approaches to evaluate the incoherent spectra.

\ \ The first and the simplest argument to prove the equivalence of the two methods is to show that the incoherent part of the spectrum calculated with the "variance" method can be transformed to a form identical to the incoherent part of the spectrum calculated with the "limit" method. To show it, we simply expand the variance spectrum into two terms
\begin{eqnarray}
S_{in}(\omega) &=&  \gamma_{1}u(\vec{r}){\rm Re}\int_{0}^{\infty} 
d\tau e^{-i(\omega -\omega_{1})\tau} \nonumber \\
&&\times  \lim_{t\rightarrow\infty}\langle \sigma_{31}(t+\tau)\sigma_{13}(t)\rangle \nonumber \\
&& -  \gamma_{1}u(\vec{r}){\rm Re}\int_{0}^{\infty} 
d\tau e^{-i(\omega -\omega_{1})\tau} \nonumber \\
&&\times  \lim_{t\rightarrow\infty}\langle \sigma_{31}(t+\tau)\rangle\langle\sigma_{13}(t)\rangle ,
\end{eqnarray}
where the first term on the right-hand side corresponds to the total spectrum. The second term corresponds to light that is elastically scattered by the atom. It is easy to see. Since in the steady-state 
$\langle \sigma_{31}(t+\tau)\rangle$ and $\langle\sigma_{13}(t)\rangle$ are independent of time, 
we can write
\begin{eqnarray}
S_{in}(\omega) =  S_{T}(\omega)  -  \gamma_{1}u(\vec{r})
\left|\langle \sigma_{31}\rangle_{s}\right|^{2} \delta (\omega -\omega_{1}) ,\label{e5}
\end{eqnarray}
where $\langle \sigma_{31}\rangle_{s}$ is the steady-state value of the average atomic dipole moment of the $|3\rangle -|1\rangle$ transition. 

\ \ We observe from Eq.~(\ref{e5}) that the resulting difference is manifestly positive since the coherent part of the spectrum cannot overweight the total spectrum. Thus, the incoherent part of the spectrum calculated with the variance method is necessarily positive. It is always positive independent of a mathematical approach adopted to evaluate the two-time correlation function of the fluctuation operators. In addition, note that the coherent part contributes only to the central component of the spectrum. Therefore, any discrepancies between the spectra observed at the Rabi sidebands do not come from the procedure of subtracting the coherent part of the spectrum.

\ \ As the second argument of the equivalence between these two methods, we consider equations of motion for the expectation values of the atomic operators and show that the solutions for the atomic two-time correlation functions appearing in Eqs.~(\ref{e2}) and (\ref{e3}) lead to the same incoherent parts of the spectrum. We consider the general case of systems whose dynamics are described by a set of inhomogeneous linear differential equations 
\begin{eqnarray}
\frac{d}{dt} \vec{X}(t) = {\bold Q}\vec{X}(t) + \vec{R} ,\label{e1}
\end{eqnarray}
where $\bold Q$ is a finite-dimensional $n\times n$ matrix of time independent coefficients, Rabi frequencies, damping rates and detunings, involved in the dynamics of a given system, $\vec{X}(t)$ is a column vector of the one-time correlation functions and $\vec{R}$ is a column vector of inhomogenous terms. The dimension $n$ of the vector $\vec{X}$ is equal to the dimension of the Hilbert space of a given system. The general case includes, as a special case, the dynamics of  $\Lambda$-type atom driven by two laser fields, considered by Xu {\it et al.}~\cite{xhy} for which $n=8$.

\ \ We first find the steady-state solution of Eq.~(\ref{e1}), which is obtained by taking the limit of $t\rightarrow \infty$, or more directly by setting the left-hand side of Eq.~(\ref{e1}) equal to zero. Thus 
\begin{eqnarray}
\vec{X}(\infty) =  -{\bold Q}^{-1}\vec{R} .\label{e6}
\end{eqnarray}

Next, we consider two-time correlation functions involved in the calculation of the spectrum of a given system. These are simply found by applying the quantum regression theorem to Eq.~(\ref{e1}) which states that for $\tau >0$ the two-time correlation functions satisfy the same equations of motion as the one-time averages~\cite{l68}. Thus,  we find that 
\begin{eqnarray}
\frac{d}{d\tau} \vec{Y}(t,\tau) = {\bold Q}\vec{Y}(t,\tau) + \langle X_{i}(t)\rangle\vec{R} ,\label{e8}
\end{eqnarray}
where $\vec{Y}(t,\tau)$ is a column vector of components 
$\langle X_{i}(t)X_{1}(t+\tau)\rangle, \langle X_{i}(t)X_{2}(t+\tau)\rangle,\ldots$.

\ \ Using the Laplace transform, we can transform Eq.~(\ref{e8}) into a set of algebraic equations, which we can solve by the matrix inversion. Thus, applying the Laplace transform, we obtain
\begin{eqnarray}
\vec{Y}(s) = \left(s{\bold I} -{\bold Q}\right)^{-1}\vec{Y}(0) 
+[ \left(s{\bold I}\right) \left(s{\bold I} -{\bold Q}\right)]^{-1}\langle X_{i}(t)\rangle \vec{R} ,\label{e9}
\end{eqnarray}
where $\vec{Y}(s)$ is a column vector composed of the Laplace transforms of the correlation functions, $s$ is a complex (Laplace transform) parameter and ${\bold I}$ is the unit diagonal matrix.

\ \ We can rewrite Eq.~(\ref{e9}) in the form
\begin{eqnarray}
\vec{Y}(s) &=& \left(s{\bold I} -{\bold Q}\right)^{-1}\left(\vec{Y}(0) 
+{\bold Q}^{-1}\langle X_{i}(t)\rangle \vec{R}\right) \nonumber \\
&& -\left(s{\bold I}\right)^{-1}{\bold Q}^{-1}\langle X_{i}(t)\rangle \vec{R} ,\label{e10}
\end{eqnarray}
which in the steady-state limit of $t\rightarrow \infty$, takes the form
\begin{eqnarray}
\vec{Y}(s) = \left(s{\bold I} -{\bold Q}\right)^{-1}\Delta \vec{Y}(0)
 +\left(s{\bold I}\right)^{-1}\langle X_{i}(\infty)\rangle\vec{X}(\infty) ,\label{e11}
\end{eqnarray}
where $\Delta \vec{Y}(0) = \vec{Y}(0) -\langle X_{i}(\infty)\rangle \vec{X}(\infty)$.

\ \ As we see from Eq.~(\ref{e11}) that the Laplace transform of the two-time correlation functions is composed of two terms, one corresponding to contributions of terms with $s\neq 0$ that reflect the presence of incoherent scattering resonances, and the other contribution from the pole at $s=0$ that reflects the coherent scattering peak. This term is proportional to delta function centered at the driving field frequency since the inverse Laplace transform of the term proportional to $\left(s{\bold I}\right)^{-1}$ gives the delta function type contribution.

\ \ We now show that the first term on the right-hand side of Eq.~(\ref{e11}) is equal to that obtained for the correlation functions of the fluctuation operators, i.e. leads to the same correlation function that is involved in the variance method of the calculation of the incoherent part of the spectrum.
This will constitute that both methods of calculating the incoherent part of the spectrum are completely equivalent.
We define a column vector of average values of the fluctuation operators $\Delta\vec{X}(t)$ whose the time evolution
is determined by the homogeneous equation of motion
\begin{eqnarray}
\frac{d}{dt}\langle\Delta \vec{X}(t)\rangle = {\bold Q}\langle \Delta\vec{X}(t)\rangle ,\label{e12}
\end{eqnarray}
from which we find, by applying the quantum regression theorem, that the vector $\Delta\vec{Y}(t,\tau)$ of the two-time correlation functions satisfies an equation of motion
\begin{eqnarray}
\frac{d}{d\tau} \Delta\vec{Y}(t,\tau) = {\bold Q}\Delta\vec{Y}(t,\tau)  ,\label{e13}
\end{eqnarray}
where the components of the vector $\Delta\vec{Y}(t,\tau)$ are 
$\langle \Delta X_{i}(t)\Delta X_{1}(t+\tau)\rangle, \langle \Delta X_{i}(t)\Delta X_{2}(t+\tau)\rangle,\ldots$.

\begin{figure}[hbp]
\includegraphics[width=\columnwidth,keepaspectratio,clip]{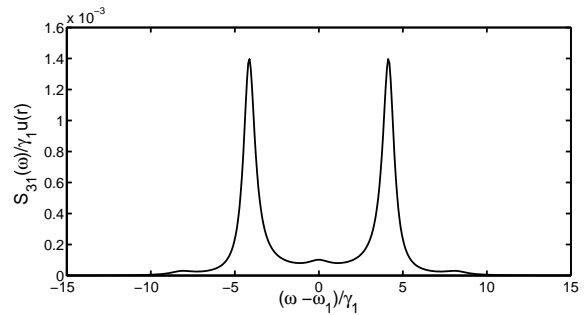}
\caption{The incoherent part of the fluorescence spectrum $S_{31}(\omega)$ calculated with 
the variance method for the same parameters as in~Fig.~2(b) of Ref.~\cite{xhy}. The spectrum is 
positive for all frequencies and exhibits Lorentzian-type structures at the Rabi sidebands.}
\label{fig1}
\end{figure}

Taking the Laplace transform of Eq.~(\ref{e13}), we find
\begin{eqnarray}
\Delta\vec{Y}(s) = \left(s{\bold I} -{\bold Q}\right)^{-1}\Delta\vec{Y}(0)  ,\label{e14}
\end{eqnarray}
which is equal to the first term on the right-hand side of Eq.~(\ref{e11}). Thus, the limit and variance methods 
both lead to the same solutions for the correlation functions involved in the definition of the incoherent part 
of the spectrum. This statement is true for an arbitrary system whose the time evolution is described by the set 
of differential equations (\ref{e1}).

\ \ Third and the last argument illustrating that there are no discrepancies between the two methods is the 
numerical analysis.
Figure~\ref{fig1} shows the incoherent part of the fluorescence spectrum $S_{31}(\omega)$ calculated numerically 
with the variance method for the same parameters as in Fig.~2(b) of Ref.~\cite{xhy}. 
Note that this is the figure where the most prominent negative values of the spectrum have been predicted. 
The negative values of the spectrum result from dispersive rather than Lorentzian-type structures at the Rabi 
sidebands.

We plot the spectrum by taking the Laplace transform of Eq.~(\ref{e13}) and solving numerically the resulting 
set of algebraic equations by matrix inversion~\cite{zf}. Evidently, the spectrum is positive for all frequencies 
and exhibits Lorentzian-type structures at the Rabi sidebands, that is the incoherent spectrum is the same as 
calculated with the limit method, shown as a dotted line 
in Fig~2(b) of Ref.~\cite{xhy}.

\ \ In summary, we have presented three simple arguments against the statement of 
Xu {\it et al.}~\cite{xhy} that the incoherent part of the fluorescence spectrum of a three-level $\Lambda$-system calculated with the variance method can have negative values. The arguments clearly show that the incoherent part of the spectrum calculated with the variance method is positive for all frequencies independent of the system considered and values of the parameters involved.


\begin{thebibliography}{99}
\bibitem{zf} A copy of the numerical program written for matlab is available from the author upon a request.
\end{thebibliography}
\end{document}